%% file: main.tex
\newcommand{\PaperTitle}{Tracking Conversations: Measuring Content and Identity Exposure on AI Chatbots}

\documentclass[10pt,sigconf,letterpaper,nonacm]{acmart}

\usepackage{array,soul}
\usepackage{makecell}
\usepackage[table]{xcolor}
\usepackage{pifont}
\usepackage{listings}
\usepackage{multirow}
\usepackage{graphicx}
\definecolor{darkgreen}{rgb}{0.0, 0.5, 0.0}

\newcommand{\para}[2][\relax]{{#1}\noindent \textbf{#2}.}

\author{Muhammad Jazlan}
\affiliation{%
    \institution{University of California, Davis}
    \city{}
    \state{}
    \country{}
}
\email{mjazlan@ucdavis.edu}

\author{Ethan Wang}
\affiliation{%
    \institution{University of California, Davis}
    \city{}
    \state{}
    \country{}
}
\email{ebwang@ucdavis.edu}

\author{Yash Vekaria}
\affiliation{%
    \institution{University of California, Davis}
    \city{}
    \state{}
    \country{}
}
\email{yvekaria@ucdavis.edu}

\author{Zubair Shafiq}
\affiliation{%
    \institution{University of California, Davis}
    \city{}
    \state{}
    \country{}
}
\email{zubair@ucdavis.edu}

\begin{document}

\title[Tracking Conversations]{\PaperTitle}

\input{0-abstract}

\maketitle

\input{1-introduction}
\input{2-background}
\input{3-methodology}
\input{4-results}

\input{5-discussion}


\bibliographystyle{ACM-Reference-Format}
\bibliography{refs}

\newpage
\input{7-appendix}

\end{document}

%% file: 0-abstract.tex
\begin{abstract}
AI chatbots are becoming a primary interface for seeking information.
As their popularity grows, chatbot providers are starting to deploy advertising and analytics.
Despite this, tracking on AI chatbots has not been systematically studied.
We present a systematic measurement of web tracking on 20 popular AI chatbots.
Under controlled settings using a sensitive prompt, we capture and compare network traffic in normal chats and, where supported, private chats.
We search for exposure of two categories of information: \textit{content}, including prompts, prompt-derived titles, chat URLs, and chat identifiers; and \textit{identity}, including names, emails, account identifiers, first-party cookies, and explicit IP/User-Agent fields in payloads.
We find that 17 of 20 chatbots share information with at least one third party.
Three chatbots share plaintext conversation text, including both prompt and response snippets, with Microsoft Clarity through session replay.
Fifteen chatbots share conversation URLs or chat identifiers with third-party advertising, analytics, or social endpoints.
Several chatbots expose user identity through support widgets, analytics, advertising, and session replay tags; in some cases, hashed emails are shared.
\end{abstract}

%% file: 1-introduction.tex
\section{Introduction}
AI chatbots have rapidly become a primary interface for seeking information, advice, and assistance.
Modern AI chatbots support open-ended conversations, answer complex queries, generate content, and adapt responses to individual users across a wide range of tasks and domains~\cite{caldarini2022literature,zhao2023survey}.
As users increasingly turn to AI chatbots for tasks previously handled by search engines or other services~\cite{techradar2025aichatbotsearch}, the web interfaces of these systems have become an important surface for privacy and safety measurements.

AI chatbot providers have growing incentives to collect and share information from these interfaces.
Advertising is a key driver: with their growing popularity, AI chatbots are emerging as an attractive channel for advertisers to reach high-intent audiences~\cite{adexchanger2025programmaticai}.
Chatbot services such as OpenAI's ChatGPT and Microsoft's Copilot have recently introduced advertising~\cite{copilotAds,openai-ads}.
Advertising creates incentives to track user activity for ad targeting (i.e., show relevant ads to a user) and conversion tracking (i.e., what a user does after seeing an ad) \cite{techcrunch2025metaadchat}.
Chatbot providers also deploy analytics, session replay, and experimentation tools to  understand and optimize usage patterns~\cite{surfshark2025chatbotprivacy,theconversation2026chatbotads}.

Tracking on AI chatbots is more consequential than tracking on many typical websites for two reasons.
First, the content can be unusually sensitive.
Users disclose health concerns, relationships, finances, and mental state to chatbots~\cite{tolsdorf2025safety, qia2026beyond, zhang2024s, malki2025hoovered}.
A traditional news site reveals what a user is reading; a chatbot, like a search engine, can reveal what a user is actively seeking, but often with more conversational detail and personal context.
Second, authentication is common.
Many chatbots allow limited use without sign-in, but reserve functionality such as chat history and file uploads for logged-in users.
This shifts tracking from browser-level identifiers, such as cookies, fingerprints, and IP addresses, toward account-level identifiers, such as email addresses and names, that are  directly linked to a person.

Despite these risks, tracking on AI chatbot websites has not been systematically studied.
Prior work has measured adjacent AI surfaces such as GPT plugins and browser-based AI assistants~\cite{wu2025depth,carrillo2026personal,vekaria2025big}, but not first-party chatbot websites, where the provider controls the interface, users are often authenticated, and conversational state is rendered directly into page metadata, URLs, and DOM content.
This creates a distinct measurement problem: traditional web tracking mechanisms now operate on pages whose state may include prompts, responses, chat identifiers, prompt-derived titles, and account identity.
It remains unclear which third parties are embedded within these platforms, what information is shared with them, and whether privacy controls such as \textit{private} chats limit third-party exposure.

To address this gap, we perform a systematic measurement of web tracking on 20 popular AI chatbots.
Under controlled settings, we capture and compare the network traffic observed using ``pregnancy test near me''---a sensitive prompt that combines health information with an implicit location request.
We analyze exposure under two categories of information: \textit{content}, including prompts, prompt-derived titles, chat URLs, and chat identifiers; and \textit{identity}, including names, emails, first-party cookies, and explicit IP/User-Agent fields in payloads.
For chatbots that support private mode, we repeat the same experiment in private chats to evaluate whether these modes reduce third-party tracking.

Our results show that tracking on AI chatbots is widespread and can expose both content and identity information.
Seventeen of 20 chatbots share information with at least one third party during a single normal chat session.
Three chatbots share plaintext conversation text, including both the user's prompt and assistant-response snippets, with Microsoft Clarity through session replay.
Fifteen chatbots share conversation URLs or chat identifiers with third-party advertising, analytics, or social endpoints.
Several chatbots expose account identity, including email addresses and names, through embedded support widgets, analytics, session replay, and advertising tags; in some cases, hashed emails are shared.
We also find that private modes substantially reduce third-party tracking.
We conclude by discussing privacy policies and privacy-enhancing design/configuration choices that reduce exposure in AI chatbots.

%% file: 2-background.tex
\section{Background and Related Works}
\label{sec:background}

\subsection{Overview of AI Chatbots}
\label{sec:background:chatbots-overview}

AI chatbots deliver personalized conversations by learning from each user's history, progressively building an understanding of their preferences, habits, and context~\cite{jiangknow, vekaria2025big}.
When a user submits a query, their prompt is combined with the provider-set system instructions and the current chat history to form the context window, and then sent as an input to a backend LLM hosted on the provider's servers.
Chatbot-as-a-service platforms (e.g., OpenRouter~\cite{openrouter}) expose multiple third-party LLMs through a single interface, forwarding queries either directly to the model provider or via the platform's own infrastructure.

The context window also includes two types of chatbot-managed memory \cite{openai2026memoryfaq}: \emph{reference memories}, inferred from past chats via the provider's proprietary memory architecture, and \emph{saved memories}, which the user has explicitly asked the chatbot to retain. Together, they let the model tailor responses to the individual.

\para{Privacy features in AI chatbots}
\label{sec:background:privacy-features}
AI chatbots offer privacy controls in three categories.
\textit{Memory controls} let users toggle reference and saved memories, and view or selectively delete saved ones~\cite{openai2026memoryfaq}.
\textit{Model improvement controls} let users opt out of having their conversations used to train or fine-tune the provider's LLMs~\cite{openai2026datacontrols}.
\textit{Private} or \textit{temporary chats}---analogous to a browser's incognito mode---isolate the session from prior memories and disable retention of the new ones~\cite{openai2026howdataisused, googleblog2025geminitemporarychats}, typically bundling the other two controls and preserving conversational utility while keeping the session private.

\subsection{Studies on Privacy in AI Chatbots}
\label{sec:background:related-works}
Prior work has examined chatbot security and privacy practices~\cite{mireshghallah2025position, ragab2024trust} and the resulting harms to users~\cite{gumusel2024user}, including LLMs' ability to infer sensitive attributes from innocuous text~\cite{staab2023beyond, fan2023well}, and adversarial configurations that elicit disclosures or exfiltrate user data~\cite{zhan2025malicious, kaya2025ai}.

Recent work has explored tracking in adjacent AI surfaces.
Wu et al.~\cite{wu2025depth} and Carrillo et al.~\cite{carrillo2026personal} audit OpenAI's GPT plugin ecosystem, finding third-party Actions that collect data prohibited by OpenAI, enable cross-GPT tracking, and disclose data practices inconsistently---only 5.8\% of Actions disclose collection in their privacy policy, and over 37\% of GPTs show broken traceability between transmissions and disclosures.
Most relevant to our work, Vekaria et al.~\cite{vekaria2025big} audit generative AI browser extensions and find them sharing webpage content, prompts, and identifiers with third-party trackers.
These works examine peripheral AI surfaces---plugins and browser extensions---rather than the first-party chatbots where users share information most directly.
We fill this gap by analyzing the network traffic of the 20 most popular AI chatbots.

%% file: 3-methodology.tex
\vspace*{-7pt}
\section{Methodology}
\label{sec:methodology}
\subsection{Data Collection on AI Chatbots}

\para{Chatbot selection}
We study web-based AI chatbots: that is, chatbot services that users access directly through the provider's website.
Our scope excludes mobile applications, browser extensions/plugins, and chatbot functionality embedded inside websites.
Our goal in this study is to study the tracking behavior of the chatbot provider's own web interface during a user-initiated conversation.
We select 20 popular AI chatbot services: ChatGPT*~\cite{chatgpt}, Gemini*~\cite{gemini}, Claude*~\cite{claude}, Grok*~\cite{grok}, DeepSeek~\cite{deepseek}, Character.AI~\cite{characterai}, Perplexity*~\cite{perplexity}, Microsoft Copilot~\cite{copilot}, PolyBuzz~\cite{polybuzz}, Kimi \cite{kimi}, Qwen Chat*~\cite{qwenchat}, Manus~\cite{manus}, Genspark~\cite{genspark}, Meta AI \cite{metaai}, Duck.ai~\cite{duckai}, SeaArt~\cite{seaart}, OpenRouter~\cite{openrouter}, Poe~\cite{poe}, Mistral*~\cite{mistral}, and ChatOn~\cite{chaton} (Services marked with * support private/temporary chats).
We use Similarweb rankings and traffic estimates as an external signal of popularity \cite{similarweb2026aichatbotsranking}.

\input{media/chatbot-data-sharing}

\para{Account setup}
For each chatbot, we created a fresh user account using either email-based registration or Google single sign-on (SSO), depending on the options supported by the chatbot service.
When the service requested profile information, we provided consistent name and account identity.

\para{Prompt}
For each chatbot, we conducted a chat session using the fixed prompt: \textit{``pregnancy test near me.''}
This prompt combines a health-related query with an implicit location request, allowing us to understand how chatbots handle potentially sensitive information.
Note that our goal is not to evaluate the quality or correctness of the chatbot's response.
Instead, the single-prompt conversation provides a controlled interaction in which any observed disclosures can be tied to the same user input across different chatbot services.

\para{Normal and private modes}
We evaluate the default \textit{normal} mode for all 20 chatbots.
For chatbots that offer a \textit{private} mode, we conduct an additional session using the same prompt.
Private chats are presented as incognito-styled conversations that claim to provide heightened privacy protections (e.g., no retention). 
This comparison allows us to measure the extent of tracking in \textit{private} vs. \textit{normal} chats.

\para{Capturing network traffic}
All testing was conducted in Chrome browser, which does not offer tracking protection at the time of writing as is offered in other browsers such as Safari's Intelligent Tracking Prevention (ITP) \cite{webkit2026trackingprevention} or Firefox's Enhanced Tracking Protection (ETP) \cite{mozilla2026enhancedtrackingprotection}.
Chrome provides a baseline for tracking that is caused by chatbot website design and third-party integration.
For each session, we began from a fresh browser state for the corresponding account and recorded the traffic generated while loading the chatbot interface, submitting the prompt, and waiting for the chatbot to complete its response.
For each session, we captured network traffic containing HTTP request and response metadata, including URLs, headers, and payloads.
Additionally, we also extract first-party cookies.
For \textit{normal} sessions, recording begins at page navigation and continues until the chatbot completes its response.
For \textit{private} sessions, recording begins when the \textit{private} session is initiated, so that traffic from initial normal-mode activity is excluded.

\vspace{-1.5mm}

\subsection{Analyzing Chatbot Traffic}
\label{subsec: analysis methodology}
We analyze each captured session in four stages: preprocessing traffic, defining search targets, matching targets against the traffic, and attributing matches to the receiving parties. 

\para[\relax]{Preprocessing}
The information transmitted by chatbot websites and embedded third-parties is often encoded or compressed before transmission.
We therefore preprocess captured URLs, headers, cookies, and payloads before searching them.
We decompress gzip- and zlib-compressed payloads when present and decode common transport encodings including base64 and URL encoding.
Decoded payloads are converted to strings so they can be searched alongside request URLs, query parameters, headers, and cookie values.

\para[\relax]{Search targets}
We search the preprocessed traffic for two broad categories of information: \textit{content} and \textit{identity}.
\textit{Content} captures what the user is asking or which conversation the user is engaged in.
It includes the user's exact prompt, readable assistant-response text when present in captured payloads, keywords drawn from the prompt (e.g., \textit{pregnancy}, \textit{test}, \textit{pregnancy test}), generated chat titles derived from the prompt, chat identifiers, and unique URLs to individual chats.
\textit{Identity} captures who the user is or how the user can be recognized across requests.
It includes the test account's first name, last name, display name, email address, internal user identifiers when observable, and any first-party cookies sent to third parties.
We also search for IP and User-Agent values when they appear as explicit fields in request payloads. 

\para[\relax]{Matching procedure}
For each search target, we look for exact string matches as well as common hashing, encoding schemes, and compression formats as used in \cite{vekaria2026understanding}.
Specifically, we generate case-variants (lowercase, uppercase), encoded forms (hex/base16, base32, base64, base64url, URL-encoded, URL-plus-encoded), compressed forms (gzip and deflate, both base64-encoded), checksums (Adler-32, CRC-32), and cryptographic hashes (MD5, SHA-1, SHA-224/256/384/ 512, SHA3-224/256/384/512, and RIPEMD-160 where available) of both the original value and each case-variant.

\para[\relax]{Party attribution}
Each request in a session is classified as first-party, if its eTLD+1 matches that of the chatbot's landing page, and third-party, otherwise.
%
%
%
Third parties are further categorized by functionality into analytics, advertising, and other.
When a domain is clearly part of the chatbot provider (e.g., openai.com on chatgpt.com), we categorize it as platform-party rather than as a third-party.


%% file: media/chatbot-data-sharing.tex
\begin{table*}[t]
    \caption{Summary of identity and content exposure on chatbots.
    For identity exposures, green=first-party, red=third-party, and orange=both; channel markers: U=url, B=body, H=header, and C=cookie.
    Content cells reflect third-party exposure only.
    For IP and User Agent, we only report exposure from the Body of the request.}
    \vspace{-2mm}
    \centering
    \small

    \renewcommand{\arraystretch}{1.2}
    \setlength{\tabcolsep}{3.5pt}

    \definecolor{cbgreen}{HTML}{009E73}
    \definecolor{cborange}{HTML}{E69F00}
    \definecolor{cbred}{HTML}{D55E00}

    \newcommand{\greencell}[1]{\cellcolor{cbgreen!25}#1}
    \newcommand{\orangecell}[1]{\cellcolor{cborange!25}#1}
    \newcommand{\redcell}[1]{\cellcolor{cbred!25}#1}

    \newcommand{\cmark}{\ding{52}}
    \newcommand{\xmark}{\ding{56}}
    \newcommand{\urlicon}{\tiny\textsf{U}}
    \newcommand{\bodyicon}{\tiny\textsf{B}}
    \newcommand{\hdricon}{\tiny\textsf{H}}
    \newcommand{\cookieicon}{\tiny\textsf{C}}

    \newcommand{\rotlabel}[1]{\rotatebox[origin=lb]{90}{\textbf{#1 }}}

    \begin{tabular}{|c|l|*{20}{>{\centering\arraybackslash}b{0.45cm}|}}
    \hline

    \multicolumn{2}{|c|}{} & 
    \rotlabel{ChatGPT} &
    \rotlabel{Gemini} &
    \rotlabel{Claude} &
    \rotlabel{Grok} &
    \rotlabel{ DeepSeek} &
    \rotlabel{Character AI} &
    \rotlabel{ Perplexity} &
    \rotlabel{ Copilot} &
    \rotlabel{ PolyBuzz} &
    \rotlabel{Kimi} &
    \rotlabel{ Qwen} &
    \rotlabel{Manus} &
    \rotlabel{ Genspark} &
    \rotlabel{Meta AI} &
    \rotlabel{Duck.ai} &
    \rotlabel{SeaArt} &
    \rotlabel{ OpenRouter} &
    \rotlabel{Poe} &
    \rotlabel{Mistral} &
    \rotlabel{ChatOn} \\

    \hline

    \multirow{3}{*}{\rotatebox{90}{\textbf{Identity  }}}
    & Email
    & \greencell{\hdricon\kern0.5pt\cookieicon}
    & \xmark
    & \redcell{\bodyicon}
    & \greencell{\bodyicon}
    & \xmark
    & \orangecell{\bodyicon}
    & \redcell{\urlicon\kern0.5pt\bodyicon\kern0.5pt\hdricon}
    & \xmark
    & \xmark
    & \xmark
    & \xmark
    & \xmark
    & \xmark
    & \xmark
    & \xmark
    & \orangecell{\bodyicon}
    & \greencell{\bodyicon}
    & \xmark
    & \redcell{\bodyicon}
    & \greencell{\cookieicon} \\
    \cline{2-22}

    & Name
    & \greencell{\hdricon\kern0.5pt\cookieicon}
    & \xmark
    & \redcell{\bodyicon}
    & \greencell{\bodyicon}
    & \xmark
    & \orangecell{\bodyicon}
    & \greencell{\bodyicon}
    & \xmark
    & \xmark
    & \xmark
    & \xmark
    & \greencell{\urlicon\kern0.5pt\hdricon\kern0.5pt\cookieicon}
    & \xmark
    & \xmark
    & \xmark
    & \greencell{\bodyicon\kern0.5pt\hdricon\kern0.5pt\cookieicon}
    & \greencell{\bodyicon}
    & \xmark
    & \redcell{\bodyicon}
    & \greencell{\hdricon\kern0.5pt\cookieicon} \\
    \cline{2-22}

    & IP and User Agent
    & \xmark & \xmark & \xmark & \xmark & \xmark
    & \greencell{\bodyicon}
    & \xmark & \xmark & \xmark & \xmark & \xmark & \xmark & \xmark & \xmark
    & \xmark & \xmark & \xmark & \xmark & \xmark & \xmark \\

    \hline

    \multirow{4}{*}{\rotatebox{90}{\textbf{Content   }}}
    & User Prompt
    & \xmark & \xmark & \xmark & \xmark & \xmark & \xmark & \xmark & \xmark & \xmark
    & \redcell{\urlicon\kern0.5pt\hdricon}
    & \xmark & \xmark
    & \redcell{\urlicon\kern0.5pt\hdricon}
    & \xmark & \xmark & \xmark & \xmark & \xmark & \xmark & \xmark \\
    \cline{2-22}

    & Prompt Keywords
    & \xmark
    & \xmark
    & \redcell{\bodyicon}
    & \xmark & \xmark & \xmark & \xmark & \xmark & \xmark
    & \redcell{\urlicon\kern0.5pt\hdricon}
    & \xmark
    & \redcell{\bodyicon}
    & \redcell{\urlicon\kern0.5pt\hdricon\kern0.5pt\bodyicon}
    & \xmark & \xmark & \xmark & \xmark & \xmark & \xmark & \xmark \\
    \cline{2-22}

    & Chat URL
    & \xmark & \xmark & \xmark & \xmark & \xmark
    & \redcell{\urlicon}
    & \xmark
    & \redcell{\urlicon}
    & \redcell{\urlicon}
    & \redcell{\urlicon\kern0.5pt\bodyicon}
    & \redcell{\urlicon\kern0.5pt\hdricon}
    & \redcell{\urlicon\kern0.5pt\bodyicon}
    & \redcell{\urlicon\kern0.5pt\hdricon\kern0.5pt\bodyicon}
    & \xmark & \xmark
    & \redcell{\urlicon\kern0.5pt\hdricon\kern0.5pt\bodyicon}
    & \xmark
    & \redcell{\urlicon}
    & \xmark
    & \redcell{\bodyicon} \\
    \cline{2-22}

    & Chat Identifier
    & \redcell{\urlicon}
    & \redcell{\bodyicon}
    & \redcell{\bodyicon}
    & \redcell{\urlicon}
    & \redcell{\bodyicon}
    & \redcell{\urlicon}
    & \xmark
    & \redcell{\urlicon}
    & \redcell{\urlicon\kern0.5pt\bodyicon}
    & \redcell{\urlicon\kern0.5pt\bodyicon}
    & \redcell{\urlicon\kern0.5pt\hdricon}
    & \redcell{\urlicon\kern0.5pt\bodyicon}
    & \redcell{\urlicon\kern0.5pt\hdricon\kern0.5pt\bodyicon}
    & \xmark & \xmark
    & \redcell{\urlicon\kern0.5pt\hdricon\kern0.5pt\bodyicon}
    & \xmark
    & \redcell{\urlicon}
    & \xmark
    & \redcell{\bodyicon} \\

    \hline
    \end{tabular}
    \label{tab:overview-normal}
    \vspace{-2mm}
\end{table*}

%% file: 4-results.tex
\section{Results}\label{sec:results}
\enlargethispage{\baselineskip}

In this section, we present the results of our measurement study.
We first characterize the third-party and platform-party flows observed during our measurements.
We then analyze exposure of the two categories of information: \textit{content} and \textit{identity}.
Finally, we compare \textit{normal} and \textit{private} chats to evaluate whether private modes reduce tracking.

\subsection{Third-Party vs. Platform-Party Flows}
\label{subsec:third-and-platform-party-tracking}
We categorize domains present on a chatbot into three categories: a \textit{first party} shares the chatbot's eTLD+1, a \textit{platform party} has a different eTLD+1 but the same owner, and a \textit{third party} has a different eTLD+1 and a different owner.

Across the 20 chatbots we test, we observe 47 unique third-party owners and 178 distinct (chatbot, third-party domain) pairs during our \textit{normal} chat sessions, after excluding each chatbot's platform-party domains.
Three chatbots contact no third parties: Gemini (\url{gemini.google.com}), Meta AI (\url{meta.ai}), and Duck.ai (\url{duck.ai}).
Instead, these services contact first- or platform-party domains.
Gemini contacts several Google-owned domains, including Google Analytics (\url{google-analytics.com}), Google Tag Manager (\url{googletagmanager.com}), and Google static content (\url{gstatic.com}).
Meta AI loads content from Facebook's CDN (\url{fbcdn.net}), while Duck.ai sends requests to \url{duckduckgo.com}.
Figure \ref{fig:normal_sankey} visualizes these data flows for advertising and analytics parties.
See Appendix~\ref{appendix:additional_traffic_flows} for data flows to all parties in normal and temporary chats.

\para{Third parties by category}
We classify each third party into one of three functional categories: advertising, analytics, and other.
Analytics and advertising appear on 17 and 12 of the 20 chatbots, respectively.
Advertising appears on fewer chatbots than analytics, but the chatbots that load advertising services often contact many advertising domains in a single session.
As a result, advertising accounts for the largest share of observed (chatbot, third-party domain) pairs: 73 of 178.
For example, SeaArt contacts 13 distinct advertisers in a single session: A8, Amazon, Google, Meta, Microsoft, Outbrain, Pinterest, Quora, Reddit, TikTok, Twitter/X, Yahoo Japan, and Yandex.
Genspark contacts a comparable number.

Note that the same owner can appear under multiple categories because organizations operate different domains and endpoints for different purposes.
Google is the clearest example: \url{doubleclick.net} and \url{googlesyndication.com} are categorized as advertising; \url{googletagmanager.com} and \url{google-analytics.com} as analytics; \url{gstatic.com}, \url{googleusercontent.com}, \url{googleapis.com}, and \url{run.app} as other.
Microsoft is similar: \url{clarity.ms} is categorized as analytics, while \url{bing.com} and \url{azure.com} are categorized as other.

\para{First-party and platform-party flows}
Some services route analytics or session-replay traffic through first- or platform-party domains.
For example, Google Analytics appears on Gemini, but we classify it as platform-party because the contacted domain is owned by Google.
Microsoft Copilot sends Microsoft Clarity traffic through the first-party endpoint \url{https://copilot.microsoft.com/cl/eus2-g/collect}.
Similarly, on ChatGPT, we observe requests to \url{chatgpt.com/ces/} whose payloads include Datadog- and Statsig-related fields.
We classify these requests as first-party because they are sent to \url{chatgpt.com}, but they remain relevant to our analysis because they show that analytics-related payloads can be routed through first-party or platform-party domains rather than through  third-party domains.

\input{media/normal_chat_sankey}

\subsection{Content Exposure}
\label{subsec:content-exposure}
We now examine exposure of \textit{content} information to different parties, observed through three mechanisms: session replay, embedded widgets, and page meta-data collected by tags and widgets loaded on the chatbot page.

\para{Session replay}
Four chatbots embed Microsoft Clarity \cite{microsoftClarity}, a session replay service.
Because the rendered DOM on a chatbot page contains the conversation itself, session replay can capture both the user's prompt and the assistant's response.

Three of the four (Genspark, SeaArt, and Chaton) transmit conversation text in plaintext to Clarity.
We observed Clar- 

\noindent ity receiving the exact prompt string (``pregnancy test near me'') on all three chatbots, along with readable portions of the assistant's reply: on Genspark, response consisted ``Here are a few pregnancy test options near you''; on SeaArt, extended roleplay response text spanned multiple sentences; and on Chaton, response text was ``Most pharmacies like CVS, Walgreens, or Rite Aid offer carry home pregnancy tests.''
Thus, Clarity receives a readable reconstruction of the conversation (i.e., prompt, response, and surrounding UI text) sufficient to infer what the user asked, what they were told, and what action did the user take.

Copilot Chat also embeds Clarity, but routes it through the first-party endpoint \url{https://copilot.microsoft.com/cl/eus2-g/collect}.
We do not observe plaintext conversation text in these payloads.
Instead, the payloads include message-specific identifiers and the order in which messages appeared in the DOM.
We therefore treat Copilot separately from the third-party plaintext session-replay exposures above.

\para{Widgets}
Genspark transmits the full prompt to Google Maps \cite{googleMaps} by embedding it within a query to the URL: \url{www.google.com/maps/embed/v1/search?q=pregnancy+test+near+me}.
The raw prompt is URL-encoded and passed to Google in parameter \texttt{q} of the map load.
This exposes the user's exact prompt to Google as part of loading the embedded map.
The map is loaded as a third-party iframe, which can attach Google cookies depending on the user's browser state---for instance, whether the user is logged into a Google account---potentially linking the prompt to the user's Google identity.

\para{Title}
Kimi, Claude, Manus, Gemini, and Genspark place the user's prompt or an extracted keyword (e.g., ``pregnancy'') into the page title, from where it is collected by tags and widgets loaded on the page.
Kimi's title reaches Google's DoubleClick via the standard \texttt{dt} parameter and is forwarded to Google Ads via a \texttt{tiba} parameter.
Claude and Manus both embed Intercom's customer-support messenger \cite{intercom}; Intercom's web widget posts page metadata (including the title) to \url{api-iam.intercom.io/messenger/web/metrics}.
Gemini sends the page title to its platform-party Google Analytics in the standard payload. 

\para{URL}
Chatbots typically assign a per-conversation URL that appears in the browser's address bar (e.g., \url{claude.ai/chat/<chat-id>}).
Fifteen of the 20 chatbots we test transmit the URL to at least one third-party through standard analytics or advertising tags, reaching 29 distinct destinations in total.
Meta Pixel's \texttt{dl} parameter, Google Analytics's \texttt{collect} endpoint, Bing's UET tag, and DoubleClick's conversion tags all collect the current page URL by default, so any chatbot that embeds these tags automatically transmits the chat identifier to the corresponding destination.
SeaArt and Genspark exhibit the broadest exposure: SeaArt transmits chat identifiers to 12 destinations, including Facebook, Google, TikTok, Yandex, Outbrain, and Quora, and Genspark to 7, including Facebook, Google, Bing, TikTok, and Microsoft Clarity.
The remaining chatbots transmit to between 1 and 4 destinations.

Unlike prompt keywords, the chat identifier does not itself reveal conversation text, but it identifies a specific conversation.
On chatbots where the chat URL is shareable or otherwise resolvable, exposing the URL can expose a pointer to the conversation; even where the URL requires authentication, the identifier still lets the receiving party associate information with the same chat.

\subsection{Identity Exposure}
\label{subsec:identity-exposure}
Next, we examine exposure of \textit{identity} information to other parties, observed through four mechanisms: widgets, analytics, advertising, and session replay.

\para{Widgets}
Embedded widgets transmit identity information as part of their own functional payloads.
On Claude and Mistral, the Intercom widget fires a ping request when the chat page loads, carrying the user's email, name, internal user ID, and user hash (Listing~\ref{lst:mistral-intercom-ping}).
The ping fires without any interaction with the support widget.
Intercom also transmits first-party cookies (\texttt{intercom-device-id-*} and \texttt{intercom- session-*}) to \url{intercom.io}.
Genspark transmits the \texttt{\_\_stripe \_mid} and \texttt{\_\_stripe\_sid} first-party cookies to \url{stripe.com}.

\para{Analytics}
Analytics and error-monitoring tags receive identity data attached to event payloads.
On Character.ai, the Sentry tag transmits session envelopes to \url{sentry.io}.
The \texttt{did} field carries the user's email, the \texttt{user\_agent} field carries the User-Agent string directly, and the \texttt{ip\_address} field is set to the sentinel \texttt{"\{\{auto\}\}"}, which instructs Sentry to record the client IP from the connection (Listing~\ref{lst:characterai-sentry}).
This differs from automatic network-layer IP exposure because the tag explicitly configures IP collection as part its payload.
Also on Character.ai, the Statsig tag transmits experimentation and feature-flag events to \url{prodregistryv2.org/v1/rgstr}.
Each event payload carries the user's email, internal user ID, IP address, and User-Agent in the \texttt{user} object alongside the experiment or feature-flag exposure being logged (Listing~\ref{lst:characterai-statsig}).
Claude transmits the \texttt{ajs\_user\_id} cookie and the \texttt{lastActiveOrg} cookie, which identifies the active organization, to Datadog.

\para{Advertising}
Advertising tags mainly receive two types of identity information: hashed emails that function as persistent cross-site identifiers \cite{ftc2024hashingnotanonymous}, and first-party cookies (in addition to third-party cookies) that function as advertising and analytics identifiers.
Perplexity transmits hashed email to Singular, and SeaArt transmits a hashed email in the \texttt{eb\_email} field to the TikTok pixel at \url{analytics.tiktok.com}.
Because the same email produces the same hash across sites, these values can support cross-site recognition \cite{ftc2024hashingnotanonymous}: Singular and TikTok can link the user's activity on these chatbot sites to activity on any other site that reports the same hashed email.
Meta's \texttt{\_fbp} cookie is the most widespread advertising identifier we observe, transmitted on Character.ai, Chaton, Genspark, Manus, PolyBuzz, and SeaArt to \url{facebook.com/tr} or \url{facebook.com/privacy_sandbox}.
Microsoft's UET cookies (\texttt{\_uetsid}, \texttt{\_uetvid}) are sent to \url{bat.bing.com} on Chaton, Copilot, Genspark, and SeaArt.
SeaArt additionally transmits TikTok's \texttt{\_ttp} and Pinterest's \texttt{\_pin\_unauth} first-party cookies to \url{tiktok.com} and \url{pinterest.com}, respectively.
Perplexity transmits the \texttt{pplx.visitor-id} cookie to Singular.

\para{Session replay}
On Chaton, Genspark, and SeaArt, the user's email and name are captured by Microsoft Clarity as part of the rendered DOM and transmitted to \url{*.clarity.ms/collect}.
Because these values are displayed in the chatbot UI (e.g., greeting messages, in-conversation mentions), they are auto-included in the session-replay snapshot as a side effect. 
Unlike the widget, analytics, and advertising examples above, this identity information is not placed into a labeled field.
It appears as readable text on the page that the replay tooling captures alongside the rest of the DOM.

\subsection{Private Chats}
\label{subsec:private-chats}
Tracking is substantially reduced in private chats.
Among chatbots with a \textit{private} mode, distinct (chatbot, third-party domain) pairs drops to 13 (shown in Figure \ref{fig:temp_sankey}), and the number of unique third-party owners drops to 3: Datadog, Mapbox, and Google.
Only ChatGPT, Claude, Grok, Perplexity and Qwen send requests to third parties.
Most importantly, we observe no \textit{identity} or \textit{content} exposure to any third party in any private chat session.

%% file: media/normal_chat_sankey.tex
\begin{figure}
    \centering
    \includegraphics[width=0.9\linewidth]{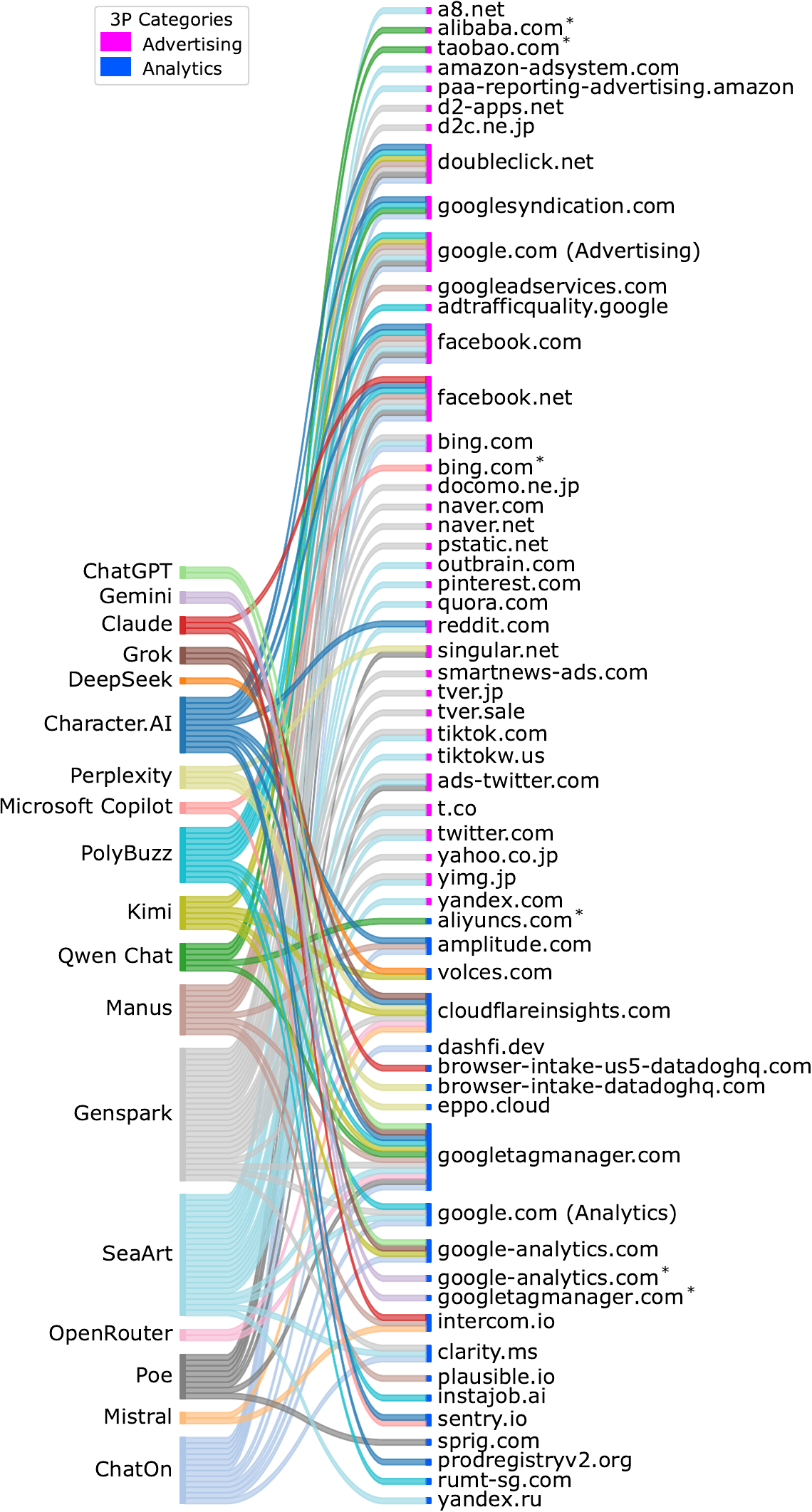}
    \vspace{-2mm}
    \caption{Sankey of data flows from chatbots to advertising and analytics third-parties in normal chats.}
    \Description[]{Sankey of data flows from chatbots to advertising and analytics third-parties in normal chats.}
    \vspace{-5mm}
    \label{fig:normal_sankey}
\end{figure}

%% file: 5-discussion.tex
\section{Conclusion \& Discussion}
\enlargethispage{\baselineskip}
We present the first systematic measurement of tracking on AI chatbots, analyzing the network traffic of the 20 most popular services.
We find that 17 of 20 chatbots embed third parties, several expose user identity to analytics and advertising vendors, and a subset transmit full conversation text in plaintext via session-replay tools.
Conversation identifiers reach a broad set of third-party destinations.
In the rest of this section, we discuss mitigations that chatbot providers can adopt, and compare the observed third-party exposure with what the chatbot privacy policies disclose to users.

\subsection{Mitigations}
Most of the exposures in our analysis result from design or configuration choices made by the chatbot provider.
In this section, we discuss mitigations that can reduce these exposures short of completely removing analytics, advertising, or embedded widgets.

\para{Page URL and title}
Placing a user's prompt, chat title, or chat identifier in the page URL or title exposes them to being tracked via embedded analytics, advertising, and support widgets that collect page metadata by default.
Providers can avoid prompt-derived text in \texttt{document.title} and redact chat identifiers from URLs reported to third-party tags.

\para{Widget integrations}
Embedded widgets should not receive the user's raw prompt unless they directly need it.
On Genspark, the Google Maps widget exposes the user's full prompt because the prompt is embedded in the Maps iframe URL as the \texttt{q} parameter.
A more privacy-preserving design would resolve the location query server-side and pass only the minimum information needed to render the map, such as coordinates or a sanitized place query.

\para{Third-party tag configurations}
Several third-party tools provide configurations that can reduce identity and content exposure.
Session-replay tools should not be deployed on authenticated conversation pages, where the rendered DOM contains both identity information and conversation text; if deployed, content masking should be employed to redact prompts, responses, and account fields.
For analytics and error monitoring, providers should avoid directly attaching identifying account fields when other identifiers suffice.
For example, the Sentry configuration on Character.AI records the client IP and transmits the user's email---neither is necessary for typical error monitoring.

\subsection{Privacy Disclosures}
We compare the observed third party data exposure with the disclosures in each chatbot's privacy policy.
We aim to assess whether the policies give users a concrete description of the third parties that receive data during chatbot use.
All 20 privacy policies \cite{openai2026privacy-policy,google2026privacy-policy,anthropic2026privacy-policy,xai2026privacy-policy,deepseek2026privacy-policy,characterai2026privacy-policy,perplexity2026privacy-policy,microsoft2026privacy-policy,polybuzz2026privacy-policy,moonshot2026privacy-policy,alibaba2026privacy-policy,manus2026privacy-policy,genspark2026privacy-policy,meta2026privacy-policy,duckduckgo2026privacy-policy,seaart2026privacy-policy,openrouter2026privacy-policy,quora2026privacy-policy,mistral2026privacy-policy,chaton2026privacy-policy} disclose that user data may be collected and used for a subset of the following purposes: analytics, advertising, service operation, personalization, safety, debugging, or support.
However, the policies vary substantially in how specifically they identify the parties involved.
Eight policies describe categories of recipients or purposes of sharing but do not name the third parties we observe in network traffic.
The gap is concerning when a privacy policy names some recipients but omits others.
Chaton, SeaArt, and Genspark each name specific third parties in their policies but omit Microsoft Clarity, despite Clarity receiving plaintext conversation text in our measurements.
In contrast, Claude and OpenRouter list a broad set of third parties present on their websites and describe the types of data shared with them.
DuckAI is a clear outlier; they do not track the user, going as far as stripping user IPs before forwarding requests to AI providers, hosted on named services.
The additional privacy disclosures shown when a user opens a private (or temporary chat) focus primarily on chat retention, model training, and personalization---not tracking \cite{openaiTemporaryChat}.
We found that private modes reduce third-party exposure in our measurements even though the disclosures do not explicitly characterize them as tracking controls.

%% file: 7-appendix.tex
\appendix

\section{Ethical Considerations}
This work studies the privacy practices of AI chatbot websites and does not rely on data collected from users.
We instead use test name, email, and account created specifically for the experimentation and is not tied to any real individual's identity.
We performed a small-scale measurement by signing up on 20 chatbot services, resulting in negligible resource utilization burden on chatbot providers: we sent a single prompt on 13 chatbots, and two prompts on the 7 chatbots that support a private or temporary chat mode.
Overall, this work does not raise any ethical concerns.

\section{Categorization of Third Parties}
We classify each third-party domain into one of three categories: \textit{advertising}, \textit{analytics}, or \textit{other}.
We first identify the eTLD+1's owner from public ownership records, then examine the owner's primary function.
A domain is classified as \textit{advertising} if its owner runs an advertising platform or serves ads on other platforms; as \textit{analytics} if not, but instead the owner provides analytics services; and else as \textit{other}.
Some owners operate distinct services across paths or subdomains of the same eTLD+1; in such cases we classify each path separately.
For example, \url{google.com/ccm} is \textit{advertising}, while \url{google.com/maps} is \textit{other}.

\section{Request Payloads}

\begin{lstlisting}[
  breaklines=true,
  basicstyle=\small\ttfamily,
  caption={Mistral \texttt{messenger/web/ping} payload to Intercom (URL-decoded, PII redacted).},
  label={lst:mistral-intercom-ping}
]
user_data={
  "email": "[REDACTED]",
  "user_id": "d34f1d9a-...",
  "user_hash": "5df3adcd5f0d...",
  "name": "[REDACTED]",
  "company": {
    "company_id": "3661535f-...",
    "name": "[REDACTED]'s Org"
  }
}
\end{lstlisting}

\begin{lstlisting}[
  breaklines=true,
  basicstyle=\small\ttfamily,
  caption={Character.ai Sentry session envelope transmitted to \texttt{sentry.io} (PII redacted). The \texttt{did} field carries the user's email, and \texttt{ip\_address: "\{\{auto\}\}"} instructs Sentry to record the client IP from the connection.},
  label={lst:characterai-sentry}
]
{
 "sid":"2017ad2c...",
 "status":"exited",
 "errors":0,
 "did":"[REDACTED]",
 "attrs":{
   "release":"e86f7f...",
   "environment":"production",
   "ip_address":"{{auto}}",
   "user_agent":"[REDACTED]"
 }
}
\end{lstlisting}

\begin{lstlisting}[
  breaklines=true,
  basicstyle=\small\ttfamily,
  caption={Character.ai Statsig event payload transmitted to \texttt{prodregistryv2.org/v1/rgstr} (PII redacted). Each event carries the user's identity alongside the experiment or feature-flag exposure being logged.},
  label={lst:characterai-statsig}
]
"user":{
 "userAgent":"[REDACTED]",
 "ip":"[REDACTED]",
 "custom":{
   "userAgeInYears":24,
   "ageCategory":"AGE_CATEGORY_O18",
 },
 "userID":"[REDACTED]",
 "email":"[REDACTED]"
}
\end{lstlisting}

\section{Additional Traffic Flows}
\label{appendix:additional_traffic_flows}

\input{media/temp_chat_sankey}
\input{media/full_normal_chat_sankey}

%% file: media/temp_chat_sankey.tex
\begin{figure}[H]
    \centering
    \includegraphics[width=1\linewidth]{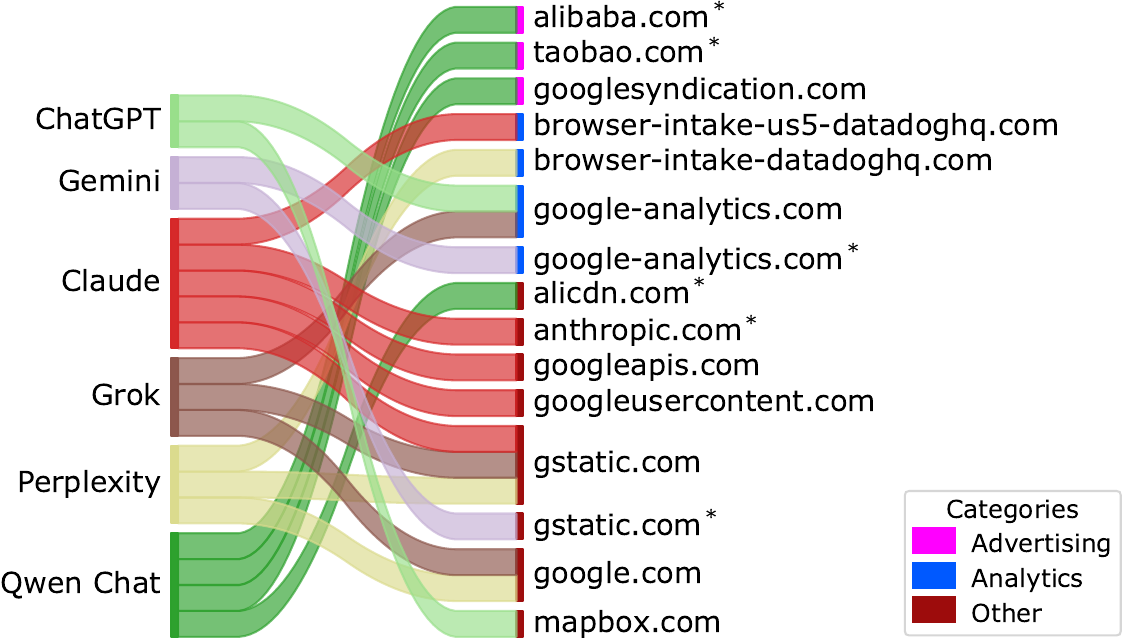}
    \caption{Sankey diagram of data flow from chatbots to all parties in temporary chats. An asterisk indicates first- or platform-party, while no asterisk means third-party.}
    \Description[]{Sankey diagram of data flow from chatbots to all parties in temporary chats.}
    \label{fig:temp_sankey}
\end{figure}

%% file: media/full_normal_chat_sankey.tex
\begin{figure}
    \centering
    \includegraphics[width=0.765\linewidth]{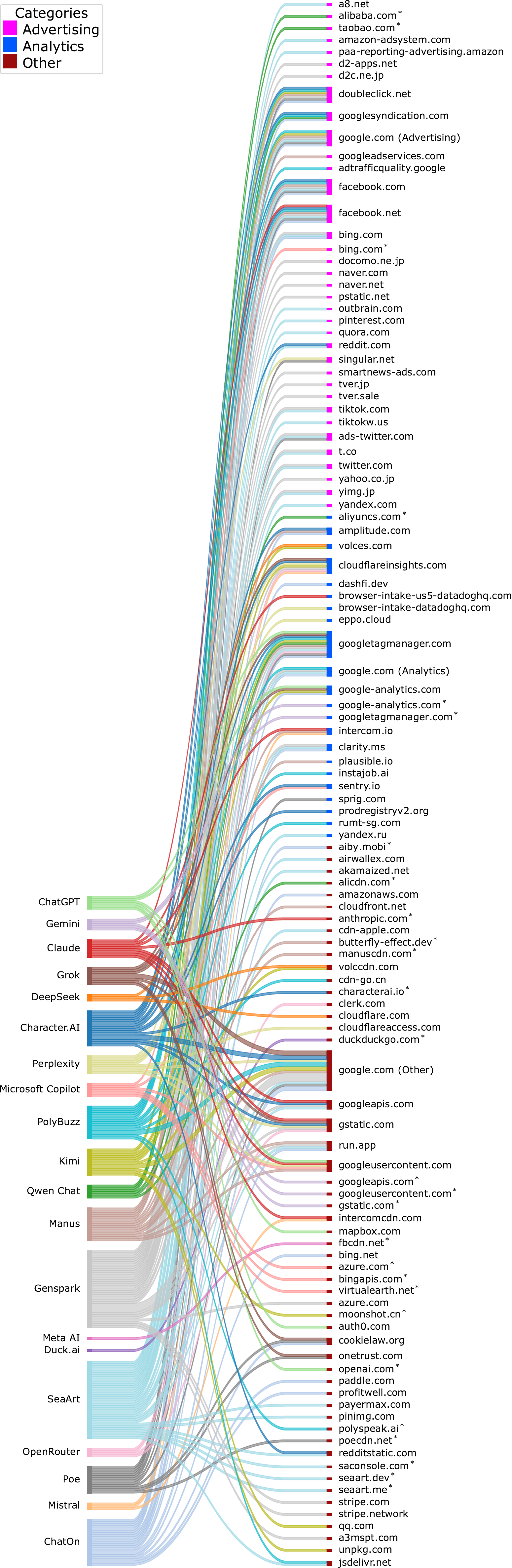}
    \caption{Sankey diagram of data flow from chatbots to all parties in normal chats. An asterisk indicates first- or platform-party, while no asterisk means third-party.}
    \Description[]{Sankey diagram of data flow from chatbots to all parties in normal chats.}
    \label{fig:normal_sankey_full}
\end{figure}